\documentclass[letterpaper, 10 pt, conference]{ieeeconf}  % Comment this line out
\usepackage{amsmath}
\usepackage{algorithm}
\usepackage[noend]{algpseudocode}
\usepackage{algorithm}
\usepackage{algpseudocode}
\usepackage{pifont}
\usepackage{amsfonts}
\usepackage{subcaption}
\usepackage{kantlipsum}       %<- For dummy text
\usepackage{mwe}  
\usepackage{float}                 %<- For dummy images
\usepackage{graphicx}
\usepackage{caption}
\usepackage{subcaption}
\usepackage{textcomp}
\usepackage[english]{babel}
\usepackage{multicol}
\usepackage[noadjust]{cite}

\usepackage{hyperref}
\hypersetup{
    colorlinks=true,
    linkcolor=blue,
    filecolor=magenta,      
    urlcolor=cyan,
}

\usepackage[noadjust]{cite}                                                        % paper

\IEEEoverridecommandlockouts                              % This command is only
\title{\LARGE \bf
Vision-Based Autonomous Driving: A Model Learning Approach
}

\author{Ali Baheri, Ilya Kolmanovsky, Anouck Girard, H. Eric Tseng, and Dimitar Filev
\thanks{Ali Baheri is with the Department of Aerospace and Mechanical Engineering, West Virginia University, Morgantown, WV 26505, USA. {\tt\small ali.baheri@mail.wvu.edu}}
\thanks{Ilya Kolmanovsky and Anouck Girard are with the Department of Aerospace Engineering, University of Michigan, Ann Arbor, MI 48109, USA. [{\tt\small ilya,anouck]@umich.edu}}
\thanks{H. Eric Tseng and Dimitar Filev are with Ford Research and Innovation Center, 2101 Village Road, Dearborn, MI 48124, USA.  [{\tt\small htseng,dfilev]@ford.com}}
}

\begin{document}

\maketitle
\thispagestyle{empty}
\pagestyle{empty}

%%%%%%%%%%%%%%%%%%%%%%%%%%%%%%%%%%%%%%%%%%%%%%%%%%%%%%%%%%%%%%%%%%%%%%%%%%%%%%%%
\begin{abstract}

We present an integrated approach for perception and control for an autonomous vehicle and demonstrate this approach in a high-fidelity urban driving simulator. Our approach first builds a model for the environment, then trains a policy exploiting the learned model to identify the action to take at each time-step. To build a model for the environment, we leverage several deep learning algorithms. To that end, first we train a variational autoencoder to encode the input image into an abstract latent representation. We then utilize a recurrent neural network to predict the latent representation of the next frame and handle temporal information. Finally, we utilize an evolutionary-based reinforcement learning algorithm to train a controller based on these latent representations to identify the action to take. We evaluate our approach in CARLA, a high-fidelity urban driving simulator, and conduct an extensive generalization study. Our results demonstrate that our approach outperforms several previously reported approaches in terms of the percentage of successfully completed episodes for a lane keeping task.

\end{abstract}

%%%%%%%%%%%%%%%%%%%%%%%%%%%%%%%%%%%%%%%%%%%%%%%%%%%%%%%%%%%%%%%%%%%%%%%%%%%%%%%%
\section{INTRODUCTION}

Autonomous driving has attracted the attention of numerous research and commercial ventures over the past decade due to its capability to change daily life and traffic. To date, several approaches have been investigated to solve autonomous driving tasks. In particular, rule-based algorithms to solve autonomous driving tasks have been developed in \cite{levinson2011towards,franke1998autonomous,muller2018driving}. Broadly speaking, these studies decompose the autonomous driving task into a few components such as perception, planning, and control and then solve each sub-task. 
%These approaches have a potential drawback that the solutions for each sub-task may not combine coherently to achieve the goal of driving.

On the other hand, Autonomous Land Vehicle in a Neural Network (ALVINN) \cite{pomerleau1989alvinn} was one of the first end-to-end learning paradigms back in 1980. It was able to learn steering angles directly from camera and laser range measurements using a neural network with a single hidden layer. Since then, significant efforts have been devoted to develop end-to-end approaches that aim to map sensor inputs to action commands from human driving data \cite{bojarski2016end,muller2006off,codevilla2018end,pan2017agile}. However, these systems cannot be generalized to unseen approaches and their performance is limited by the coverage of previously experienced human driving data. 
\begin{figure}[H]
    \centering
    \includegraphics[width=.45\textwidth]{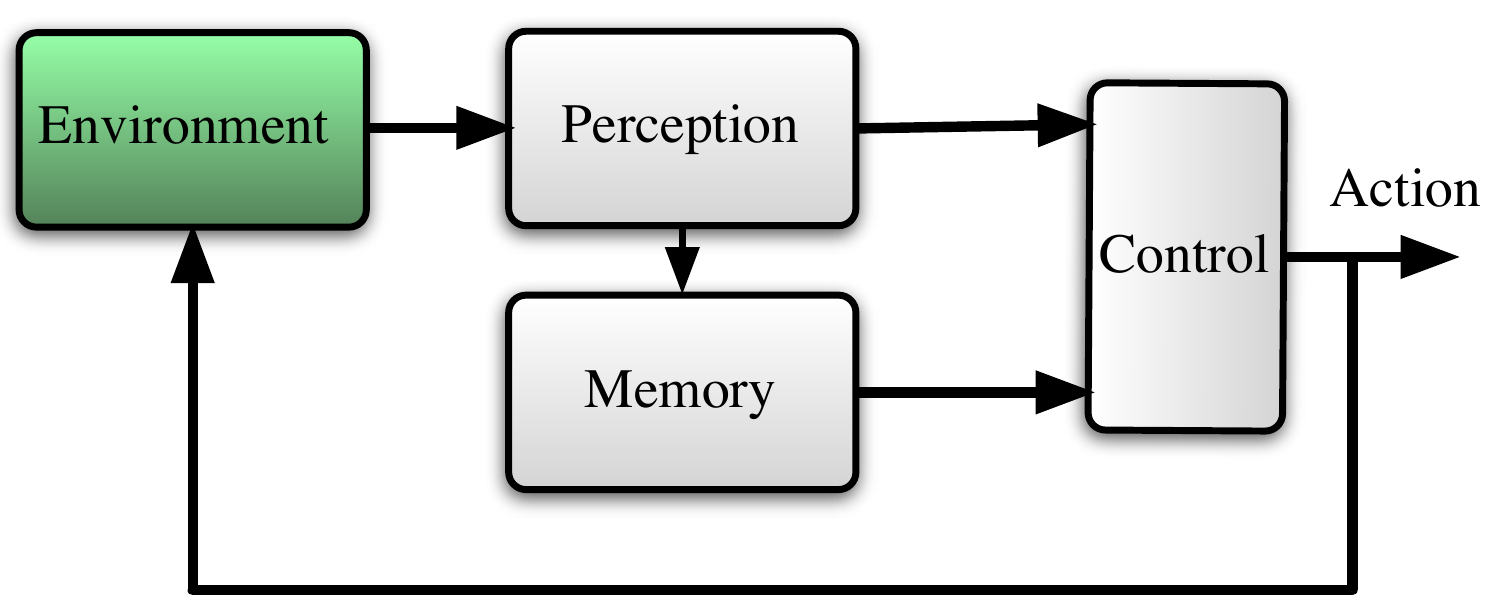}
    \caption{High-level overview of the proposed framework. We jointly consider the perception and control problems to solve the lane keeping task for an autonomous vehicle in a high-fidelity urban environment.}
    \label{fig:overall_framework}
\end{figure}
Reinforcement learning is a branch of AI in which the agent interacts with its environment, aiming to maximize a reward function. Reinforcement learning has shown dramatic success at learning to play Atari games and acting in complex environments to gain human-level control \cite{mnih2015human,silver2016mastering,silver2018general}. 

Reinforcement learning algorithms can be either model-free or model-based. Several existing studies have focused on utilizing model-free reinforcement learning for autonomous driving that aims to learn a policy without explicitly modeling the environment. \cite{liang2018cirl,shalev2016safe}. 
%In general, model-free reinforcement learning algorithms tend to be simple and slow, particularly in real-world scenarios.
There exists an extensive body of literature on training a policy by learning a model of the environment. For instance, \cite{deisenroth2011pilco} learns a probabilistic dynamics model and then exploits this model to develop a policy. 
%Furthermore, recurrent world models, introduced in \cite{ha2018world}, use an internal world model to train an agent entirely in a simulated environment, and transfer the learned policy effectively to the real-world.

Inspired by how humans perceive and make decisions in the world, the authors of \cite{ha2018world} propose an approach in which a model of the environment is learned using a combination of deep learning algorithms. Based on the learned model, a policy is trained to identify the action to take at each time-step. Specifically, they introduced recurrent world models, used as an internal world model to train an agent entirely in a simulated environment, and transferred the learned policy effectively to the real-world.

To-date, the majority of research in autonomous driving has focused on model-free reinforcement learning approaches \cite{mirchevska2018high,kendall2018learning,baheri2019deep}. Comparatively, fewer studies have focused on the impact of learning a model of the environment and training a policy based on it.
In this work, we take inspiration from \cite{ha2018world}, build a recurrent model that is able to mimic the complex urban driving environment, and study the impact of exploiting the learned model during policy training. The model consists of the following key components: a perception module, a memory module, and a controller. The perception module is a variational autoencoder (VAE), which reduces the dimensionality of the data and compresses incoming observations into a latent state representation. The memory module is a recurrent neural network (RNN) intending to predict the latent state representation of the next frame given the current frame and current action. Finally, a controller utilizes these latent representations and learns to take action.

Our key contributions can be summarized as: 

\begin{itemize}

\item We demonstrate a model-based reinforcement learning methodology for autonomous driving in which a model of complex driving environment is built and utilized for training. 

\item We show that by learning a model of the autonomous driving environment and then training a policy based on it, it is possible to significantly improve the sample efficiency and achieve lane keeping for an autonomous vehicle.

\item We evaluate our proposed approach in CARLA, a high-fidelity urban driving simulator, and conduct a generalization study to demonstrate the robustness of the approach in varying conditions.

\end{itemize}

The remainder of this paper is structured as follows: In Section \ref{ref:problem statement}, we introduce the problem at hand and formalize the system architecture. Section \ref{sec:method} introduces our approach, including VAE and RNN. Section \ref{sec:results} provides detailed results, demonstrating improvement over the baseline approaches.

\section{Problem Statement and System Architecture}
\label{ref:problem statement}

The objective considered in this paper is to train an agent for lane keeping in an urban driving high-fidelity simulator. We initialize the agent somewhere in a town and it has to reach a goal before the end of the episode. The episode is considered as successful when the vehicle reaches the goal. Otherwise, the episode terminates when the vehicle collides with an obstacle, or when a time budget is exhausted. 

We formalize the problem as a Markov decision process (MDP) where at each time-step $t$, the agent interacts with the environment, receives the state $s_t \in \mathcal{S}$, and performs an action $a_t \in \mathcal{A}$. As a result, the agent receives a reward $r_t \in \mathcal{R}$ and ends up in a new state $s_{t+1}$. The goal is to find a policy, $\pi$, that maps each state to an action with the goal of maximizing expected cumulative reward.

\begin{figure*}[t]
    \centering
    \includegraphics[width=.75\textwidth]{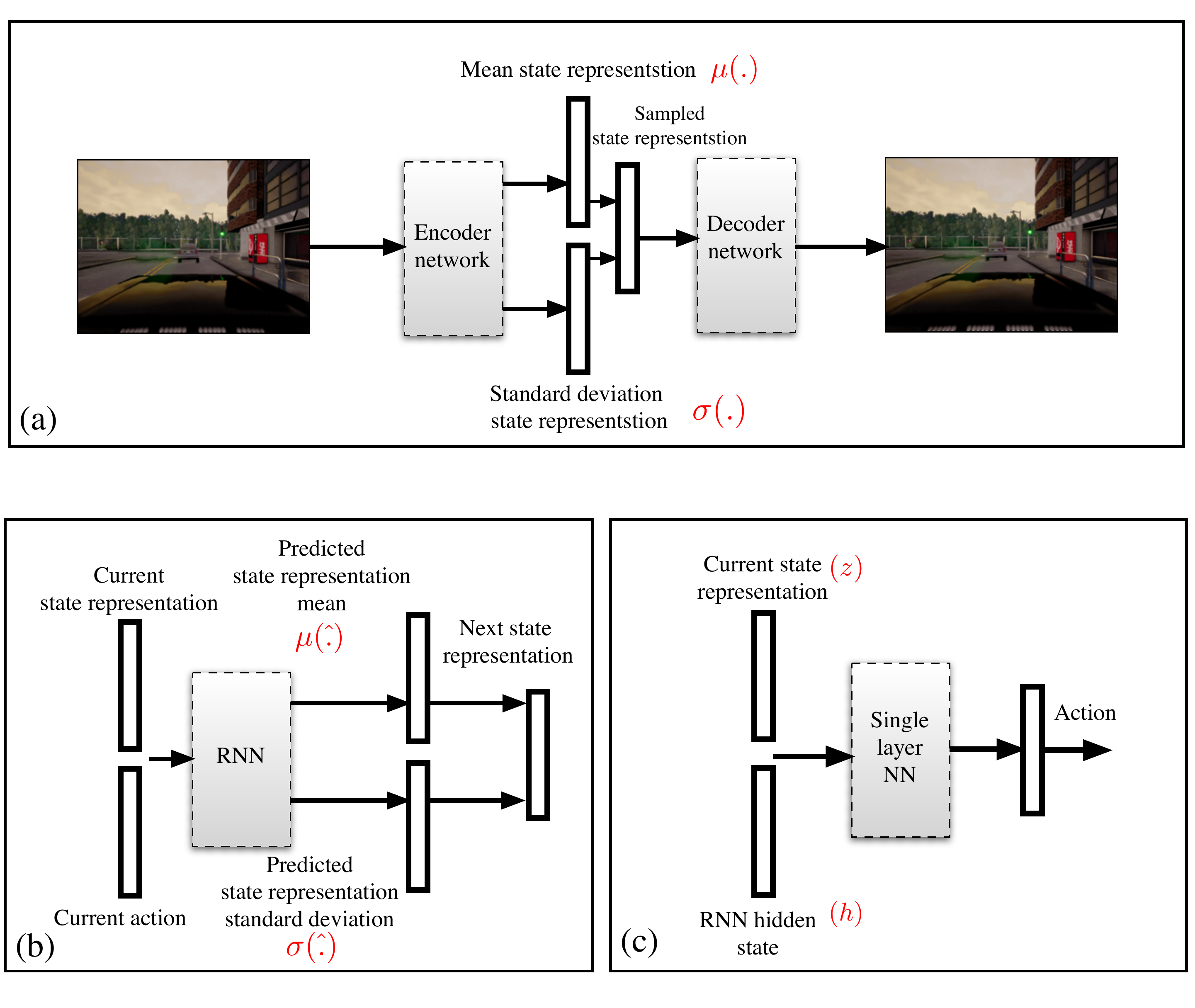}
    \caption{Our proposed architecture. (a) First, we train a VAE which encodes raw RGB observations into a compressed state representation (Sec. \ref{sec:vae}). (b) Next, we train a RNN to predict the latent state representation of the next frame (Sec. \ref{sec:rnn}). (c) Finally, a controller maps current latent state representation and RNN hidden states from previous time-step to action (Sec. \ref{sec:controller}).}
    \label{fig:framework_details}
\end{figure*}

\subsection{State space}

We focus on vision-based autonomous driving. At each time-step $t$, the agent receives a single image from a front-facing camera.

\subsection{Action space}

We consider a continuous action space in this work. The action space consists of steering angle, acceleration, and brake. The steering angle is represented by a real number between $-1$ and $1$. The acceleration and brake are represented as real numbers between $0$ and $1$. We assume that acceleration and braking do not occur at the same time. Therefore, we combine the acceleration and brake commands into a single value between $-1$ to $+1$, where the values between $-1$ and $0$ correspond to the brake command and the values between $0$ and $1$ correspond to the acceleration command.

\subsection{Reward function}

We take inspiration from the definition of the reward function in \cite{Dosovitskiy17} and define the reward as follows:

\begin{equation}
r = k_1 (d_{t} - d_{t-1}) + k_2 (v_{t} - v_{t-1}) -k_3 (s_{t} - s_{t-1}) - k_4 (o_{t} - o_{t-1}),
\label{eqn:reward}
\end{equation}
where the reward consists of sum of four terms: distance traveled towards the goal $d$ in km, speed $v$ in $\frac{km}{h}$, intersection of the agent with the sidewalk $s$, and with the opposite lane $o$.

%%%%%%%%%%%%%%%%%%%%%%%%%%%%%%%%%%%%%%%%%%%%%%%%%%

% Methodology

%%%%%%%%%%%%%%%%%%%%%%%%%%%%%%%%%%%%%%%%%%%%%%%%%%
\section{Methodology}
\label{sec:method}

We address the problem of autonomous urban driving based on a monocular camera feed. The
overall architecture of the proposed driving system is illustrated in Figure 1. The system consists of
three components: a perception module, a memory module, and a controller module. The perception
module takes as input a raw RGB image and outputs a low-level latent representation of the raw image. The memory module then takes a vector of latent variables and the current action as input and produces the predicted latent representation for the next frame. Given the low-level representation of the current frame and hidden states from RNN, the controller generates the following actions: steering angle, acceleration, and brake. Algorithm \ref{ref:alg1} summarizes this procedure. We describe each of the three modules in detail in the next subsections.

\begin{algorithm}{}
\caption{}
\label{alg:euclid}
\begin{algorithmic}[1]
%\Procedure{}{}
\State Collect several trajectories from a policy or expert driver to constitute a model of the environment
\State Train VAE to encode frames into a compressed latent representation 
\State Train RNN to predict the latent representation of the next frame
\State Train controller to maximize expected cumulative reward using evolutionary algorithms
%\EndProcedure
\end{algorithmic}
\label{ref:alg1}
\end{algorithm}

\subsection{VAE model}
\label{sec:vae}
The perception module is a VAE that builds the abstract representation of the environment and is used to compress the visual information the agent receives at each time-step from the environment. 

A VAE consists of three main components: an encoder, a decoder, and a loss function. The encoder, $\phi_{\mathrm{encode}}(.)$, is a neural network. Its input is a raw RGB image, $s$, and its output is a latent state representation $z$.
%
%\begin{equation}
%\mu({s}_{t}),\sigma({s}_{t}) \leftarrow  \phi_{\mathrm{encode}}(s_t),
%\end{equation}
%
The latent state representation is sampled from,

\begin{equation}
z_t \sim \mathcal{N}\Big(\mu(s_t),\sigma(s_t)\Big),
\end{equation}
where $\mu(.)$ and $\sigma(.)$ represent the mean and standard deviation of state representation, respectively. The decoder, $\phi_{\mathrm{decode}}(.)$, is another neural network. It takes the representation $z$ and outputs the reconstructed image,

\begin{equation}
\hat{s}_t = \phi_{\mathrm{decode}}(z_t).
\end{equation}
We define the loss function for the VAE as follows:

%the negative log-likelihood with a regularizer \cite{doersch2016tutorial},

\begin{equation}
%l(\theta,\phi) = -\mathbf{E}_{z\sim q_\theta(z|s)}[\mathrm{log}p_\phi(s|z)] + KL\big(q_\theta(z|s) || p(z)\big)
L_{\mathrm{VAE}} = \mathbb{KL}\Big(\mathcal{N}\big(\mu(s_t),\sigma(s_t)\big)\parallel \mathcal{N}(0,I)\Big) + \left \| s_t-\hat{s}_t \right \|^2.
\end{equation}
The first term is the Kullback-Leibler divergence rate (or relative entropy), which is a measure of how the predicted frame diverges from the original frame when considered as a probability distribution. The second term is the reconstruction loss measuring how close the predicted frame is to the original frame. The loss function is optimized using gradient descent with respect to the parameters of the encoder and decoder networks.

\subsection{RNN model}
\label{sec:rnn}

The memory module is a predictive model of what the VAE is supposed to predict one time-step ahead. Given a compressed representation $z$ from VAE, the memory module is used to model the dynamics and predict the next $z$. We specifically use a Long Short-Term Memory (LSTM) network \cite{gers1999learning}. 
%The vector of hidden states is represented by $h$. 
At each instant, the LSTM model, captures a latent understanding of the current state of the agent to predict the next $z$, one time-step ahead, based on the previous $z$ and the previous action, with its internal hidden state $h$. In other words, the LSTM model takes the encoded image data from the VAE and actions as inputs and returns one time-step ahead encoded image data from the VAE as an output. To encode temporal information, the LSTM models the environment dynamics in the latent state space. The predicted latent state representation is sampled from,

%\begin{equation}
%\mu(\hat{s}_{t+1}),\sigma(\hat{s}_{t+1}) \leftarrow  g(z_t,a_t),
%\end{equation}

\begin{equation}
\hat{z}_{t+1} \sim \mathcal{N}\Big(\mu(\hat{s}_{t+1}),\sigma(\hat{s}_{t+1})\Big).
\end{equation}
Here, $\mu(\hat{.})$ and $\sigma(\hat{.})$ represent the predicted mean and standard deviation of the state representation, respectively.

We train LSTM to minimize the Kullback-Leibler divergence rate between the predicted latent representation and actual latent representation from VAE,

\begin{equation}
L_{\mathrm{LSTM}} = \mathbb{KL}\Big(\mathcal{N}\big(\mu(\hat{s}_{t+1}),\sigma(\hat{s}_{t+1})\big) \parallel \mathcal{N}\big(\mu({s}_{t+1}),\sigma({s}_{t+1})\big)\Big).
\end{equation}

\subsection{Controller model}
\label{sec:controller}

The controller is a fully connected neural network with a single layer that takes the concatenation of the current latent representation of the frame, $z$, and the hidden state of the LSTM as the inputs and returns three outputs neurons corresponding to the three actions,

\begin{equation}
a_t = \textbf{W} \begin{Bmatrix}
z_t & h_{t-1} 
\end{Bmatrix}  + \textbf{b},
\label{eqn:controller}
\end{equation}
%\begin{equation}
%a_t \sim c(z_t,h_{t-1})
%\end{equation}
%which takes both the latent encoding of the current frame, and the hidden state of the MDN-RNN given past latents and actions as input and outputs an action. 
%
where $\textbf{W}$ and $\textbf{b}$ are the weight matrix and bias vector of the neural network, respectively. $z_t$ is the latent state representation at current time-step from the VAE model and $h_{t-1}$ is the hidden states at previous time-step from the RNN model. 

We train the controller to maximize the expected cumulated reward. To achieve this goal, we use a form of reinforcement learning that utilizes an evolutionary algorithm known as the Covariance Matrix Adaptation Evolution Strategy (CMA-ES) \cite{hansen1996adapting}. CMA-ES is an evolutionary algorithm for solving nonlinear, non-convex, and black-box optimization problems in the continuous domain. 
%In CMA-ES population is represented by a full covariance multivariate Gaussian. 
Compared to traditional reinforcement learning algorithms, the benefit of utilizing evolutionary algorithms  is that only the final cumulative reward is given to the optimizer, instead of the entire history. Recent work has shown that evolution strategies are a promising alternative to traditional reinforcement learning algorithms \cite{salimans2017evolution}. 

The ultimate goal of CMA-ES is to solve for \textbf{W} and \textbf{b} in Eq. $\ref{eqn:controller}$ such that the cumulative reward (Eq. $\ref{eqn:reward}$) is maximized. Algorithm \ref{ref:alg2} summarizes this procedure. For the problem at hand, in line $1$, the reward function is evaluated on multiple randomly initialized populations of decision variables, \textbf{W} and \textbf{b}. The best $20\%$ of the population is then sorted in line $2$. Next, based on the best solution and the mean of the current generation, the covariance matrix of the next generation is computed in line $3$. Finally, a new set of decision variables is sampled using the updated mean and updated covariance matrix in line $4$.

%Inspired by natural evolution, evolution strategies belongs to black box optimization algorithms. Particularly, at every iteration, a population of parameter vectors is perturbed and their performance metric is evaluated. The highest scoring parameter vectors are then recombined to form the population for the next generation. This procedure is iterated until the performance metric is optimized. 

%Algorithms in this class differ in how they represent the population and how they perform mutation and recombination \cite{coello2007evolutionary}. 

%CMA-ES is an evolutionary algorithm for solving nonlinear and non-convex black-box optimization problems in continuous domain. In CMA-ES population is represented by a full covariance multivariate Gaussian. CMA-ES has been successful in solving optimization problems in low to medium dimensions. 
%

\begin{algorithm}{}
\caption{Covariance Matrix Adaptation Evolution Strategy (CMA-ES)}
\label{alg:euclid}
\begin{algorithmic}[1]
%\Procedure{}{}
\State Compute the performance metric of each candidate solution in each generation
\State Isolate the best $20\%$ of the population in generation
\State Select the best solutions with the mean of the current generation and compute the covariance matrix of the next generation
\State Sample a new set of candidate solutions using the updated mean and updated covariance matrix
%\EndProcedure
\end{algorithmic}
\label{ref:alg2}
\end{algorithm}

\section{Results}
\label{sec:results}

\begin{figure}[t]
    \centering
    \includegraphics[width=0.45\textwidth]{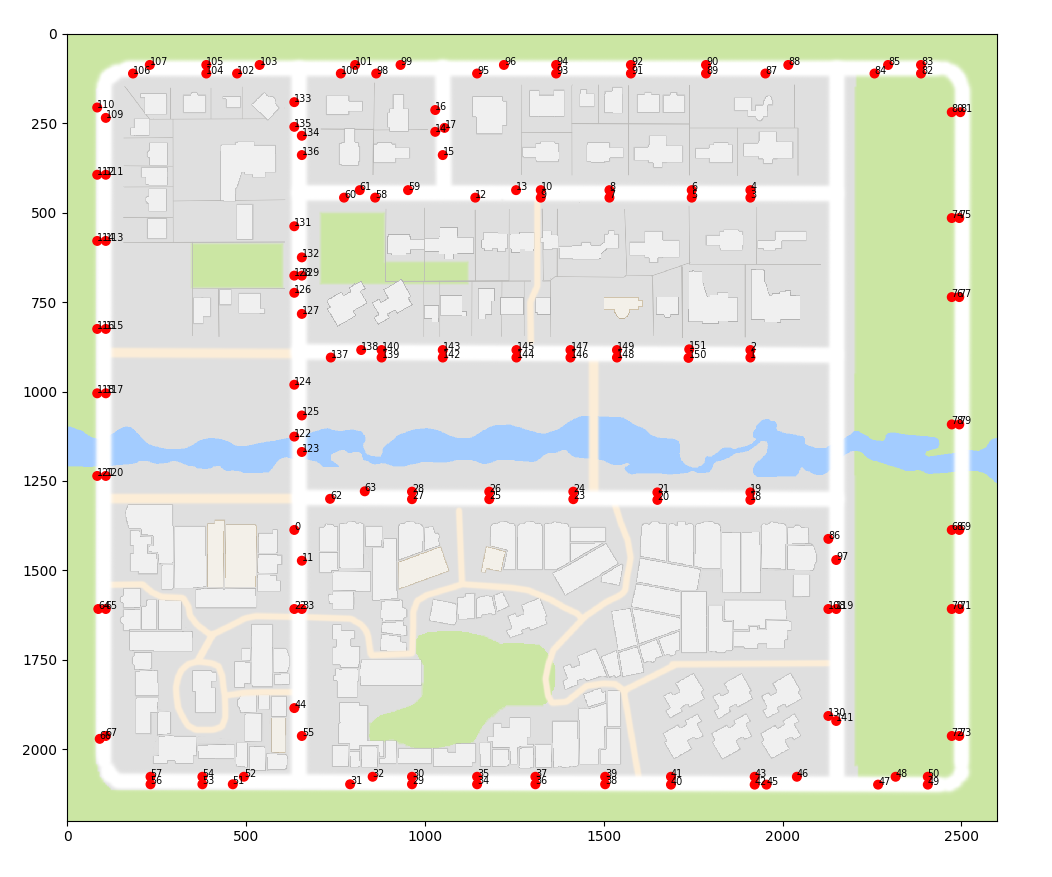}
    \caption{The map for the town that is used for testing in our experiments \cite{Dosovitskiy17}. The agent is initialized somewhere in a town and has to reach a specified goal.}
    \label{fig:position_CARLA}
\end{figure}

To evaluate the proposed approach we use CARLA, an open-source simulator for urban driving \cite{Dosovitskiy17}. Two CARLA towns used in our experiments are illustrated in Fig. 4. For fair comparison with results reported in \cite{Dosovitskiy17}, we use Town $1$ for training and Town $2$ exclusively for testing. The two towns differ in their size, layout, and style. Furthermore, CARLA provides multiple environmental conditions which can be used for further generalization study.

\subsection{Implementation settings}

\subsubsection{Data}

To build a model of the environment (Step $1$ in Algorithm \ref{ref:alg1}), we collect training data in simulation, using the CARLA platform. The training dataset comprises $5$ hours of human driving in Town $1$. Our dataset includes RGB images from a front-facing camera along with associated actions.

\subsubsection{Network architectures}

We resize image frames into $(128, 128, 3)$ pixels to feed into the encoder network. The encoder and decoder network in the VAE model comprise four convolutional layers.
%with $32, 64, 128, 256$ channels, a flatten layer, and a linear layer of $128$ units.
%
%The convolutional layers use $4 \times 4$, $4 \times 4$, $3 \times 3$, $3 \times 3$ filter size and strides of $2, 2, 1, 1$. The decoder network reverses the operations. 
%
%except the last deconvolutional layer uses $6 \times 6$ filter size. 
%
The convolutional and deconvolutional layer use the $\mathrm{Relu}$ activation function. 
%The last convolutional layer uses $\mathrm{tanh}$ activation. 
The latent state dimension of VAE is set to $128$. Furthermore, we use one layer LSTM of $256$ hidden size to encode temporal information. The controller network comprises one single dense layer neural network. 
%We implemented our network training in Tensorflow on a PC-based workstation.

\begin{figure*}[t]
   \centering
   \includegraphics[width=1\textwidth]{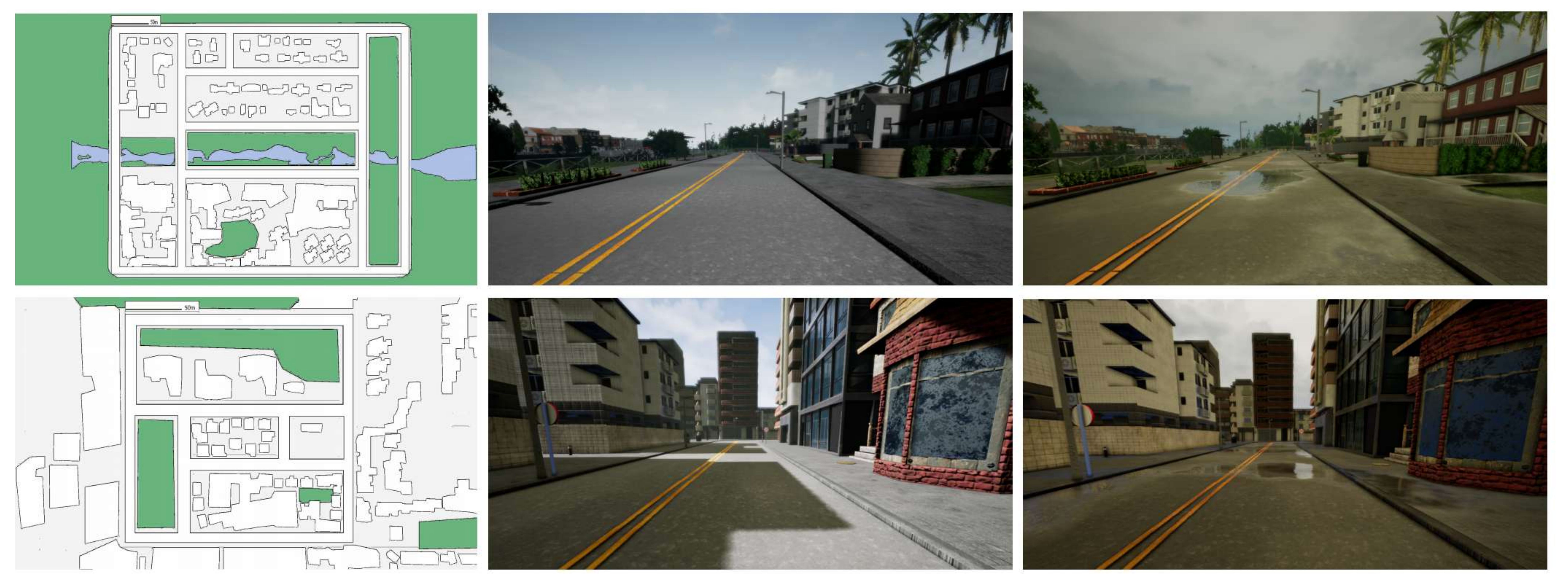}
   \caption{CARLA supports two towns and different weather conditions. Town $1$ is used for training and town $2$ is used exclusively for testing. This figure illustrates those two towns and example scenes with different weather conditions. We use unseen town and weather conditions for our generalization study.}
   \label{fig:test}
\end{figure*}
\subsection{Comparison with baselines}

We evaluate the performance of our proposed approach and compare it with three other approaches which have been previously reported using the CARLA benchmark:
\begin{itemize}
\item \textbf{Approach 1} (Ours): A VAE encodes the input image into a compressed state representation. A LSTM predicts the latent representation of the next frame based on the current latent representation. Finally, the controller determines the action to take at each time-step based on LSTM hidden states and current latent state representation.

\item \textbf{Approach 2} (MP): A modular pipeline decomposes the driving task into perception, planning, and control. For the perception module, semantic segmentation is used to identify lanes and road limits. Furthermore, a local planner utilizes a rule-based state machine to find predefined policies. Eventually, a PID controller is used to actuate the steering, acceleration, and brake.

\item \textbf{Approach 3} (IL): Imitation learning utilizes a dataset recorded by expert drivers in the training town. Particularly, it takes the images and directly trains the model via supervised learning using human driving data.

\item \textbf{Approach 4} (RL): Reinforcement learning utilizes the asynchronous advantage actor-critic (A3C) algorithm. Multiple simulation threads are run in parallel in the asynchronous reinforcement learning method.
\end{itemize}

\begin{table*}[t]
\centering
\small{
\begin{tabular}{lcccc}
\hline
\multicolumn{1}{c}{Task} & \begin{tabular}[c]{@{}c@{}}MP\\ (new town)\end{tabular} & \begin{tabular}[c]{@{}c@{}}IL\\ (new town)\end{tabular} & \begin{tabular}[c]{@{}c@{}}RL\\ (new town)\end{tabular} & \begin{tabular}[c]{@{}c@{}}Ours\\ (new town)\end{tabular} \\ \hline
\textbf{Straight} & 92 & 97 & 74 & \textbf{98} \\ \hline
\end{tabular}}
\caption{Quantitative comparison of our method with other autonomous driving approaches after training for the straight task. This table demonstrates the percentage of successfully completed episodes.}
\label{my-label}
\end{table*}
%%%
\begin{table*}[h]
\centering
\small{\begin{tabular}{lcccc}
\hline
\multicolumn{1}{c}{Task} & \begin{tabular}[c]{@{}c@{}}MP\\ (new weather)\end{tabular} & \begin{tabular}[c]{@{}c@{}}IL\\ (new weather)\end{tabular} & \begin{tabular}[c]{@{}c@{}}RL\\ (new weather)\end{tabular} & \begin{tabular}[c]{@{}c@{}}Ours\\ (new weather)\end{tabular} \\ \hline
\textbf{Straight} & 100 & 98 & 86 & \textbf{100} \\ \hline
\end{tabular}}
\caption{Generalization study. CARLA supports several weather conditions such as cloudy noon, mid rainy noon, wet cloudy sunset, hard rain sunset, etc. We change the weather to an unseen weather and report the percentage of successfully completed episodes for each approach.}
\end{table*}

\begin{table*}[h]
\centering
\small{\begin{tabular}{lcccc}
\hline
\multicolumn{1}{c}{Task} & \begin{tabular}[c]{@{}c@{}}MP\\ (new town/weather)\end{tabular} & \begin{tabular}[c]{@{}c@{}}IL\\ (new town/weather)\end{tabular} & \begin{tabular}[c]{@{}c@{}}RL\\ (new town/weather)\end{tabular} & \begin{tabular}[c]{@{}c@{}}Ours\\ (new town/weather)\end{tabular} \\ \hline
\textbf{Straight} & 50 & 80 & 68 & \textbf{88} \\ \hline
\end{tabular}}
\caption{We change the town and weather to an unseen weather and town and report the percentage of successfully completed episodes for each approach.}
\end{table*}

After training, we select $50$ start-goal points in Town $2$, conduct experiments for each pair and report the results. The results for MP, IL, and RL have been reported in \cite{Dosovitskiy17}.
Table 1 shows the percentage of successfully completed episodes in each approach for the straight driving task. One can conclude that the proposed approach outperforms baseline approaches. Furthermore, it is of interest to study to what extent a machine learning algorithm can be generalized. To achieve this goal, Table 2 and Table 3 present the generalization study results where the weather condition and the town have been changed to an unseen scenario.
%A demo of the results can be seen here (The link for videos coming soon).

\subsection{Discussion}

One of the main benefits of model-based reinforcement learning algorithms is sample efficiency. We solved the problem at hand with significantly smaller amounts of training data. Specifically, while $5$ hours of driving data in the simulator is used in our approach to solve the task, $12$ days and $14$ hours of driving data are needed to solve the same problem using pure reinforcement learning (Approach $4$) and imitation learning (Approach $3$), respectively. However, the main limitation of our approach is to what extent we can learn a model of the environment. In other words, the performance of our approach heavily depends on the quality of the model. In general, learning a perfect model of the environment remains a challenging task, particularly in real-world complex environments.

Another limitation of our approach resides in the perception aspect. In this work, we used the variational autoencoder to encode the input RGB image into a low-level representation of the image. However, the ability of this low-level representation to solve complex driving tasks is limited. In future work, we combine this low-level representation from VAE with a mid-level representation such as affordance indicators \cite{chen2015deepdriving} which represent state variables based on direct perception with physical meaning.

 %%%%%%%%%%%%%%%%%%%%%%%%%%%%%%%%%%%%%%%%%%%%%%%%%%%%%%%%%%%%%%%%%%%%%%%%%%%%%%%%%%%%%%%%%%%%%%%%%%%%

\section{CONCLUSIONS}

In this paper, we proposed an approach to the integrated treatment of the perception and control and demonstrated this approach for lane keeping in a high-fidelity urban driving simulator. Our approach relied on building a model of the environment, then training a policy exploiting the learned model to identify the action to take at each time-step. To build a model for the environment, we first trained a variational autoencoder to encode the input image into a compressed latent representation. We then utilized a recurrent neural network to predict the latent representation of the next frame. Finally, we trained a controller based on the latent representations at hand to determine the action to take by leveraging an evolutionary-based reinforcement learning algorithm. We evaluated our approach in CARLA  and conducted generalization studies. The results suggest that our approach is capable of outperforming existing approaches in terms of the percentage of successfully completed episodes for lane keeping tasks.

\section{Acknowledgment}

This work is supported by Ford Motor Company.

\newpage
\bibliographystyle{unsrt}
\bibliography{Bo_Summer2016}

\end{document}